\documentclass{article}

\usepackage{arxiv}
\usepackage[utf8]{inputenc} 
\usepackage[T1]{fontenc}    
\usepackage{hyperref}       
\usepackage{url}            
\usepackage{booktabs}       
\usepackage{amsfonts}       
\usepackage{nicefrac}       
\usepackage{microtype}      
\usepackage{lipsum}
\usepackage{indentfirst}
\usepackage{setspace}

\usepackage{graphics}      
\usepackage{graphicx}      
\usepackage{longtable}     
\usepackage{float}

\newcommand{\beq}{\begin{equation}}
\newcommand{\eeq}{\end{equation}}
\newcommand{\bea}{\begin{eqnarray}}
\newcommand{\eea}{\end{eqnarray}}

\title{CASIMIR EFFECT IN A SCHWARZSCHILD-LIKE WORMHOLE SPACETIME}

\author{{\Large
  A. C. L. Santos,\thanks{alana.santos@aluno.uece.br}\,\,\,\, C. R. Muniz\thanks{celio.muniz@uece.br}\,\,\,\, and\,\,\, L. T. Oliveira\thanks{leonardo.tavares@uece.br}}\\
{\Large Universidade Estadual do Cear\'{a}}\\
{\Large Faculdade de Educa\c{c}\~{a}o, Ci\^{e}ncias e Letras de Iguatu}\\
{\Large Av. Dario Rabelo, s/n,
CEP: 63.500-000, Iguatu, CE, Brazil}\\
}

\begin{document}
\maketitle
\vspace{1.0cm}
\doublespacing 
\begin{abstract}
{\large 
In this paper, we investigate the role of gravito-inertial effects on the Casimir energy of a massless scalar field confined between two parallel plates orbiting a static and zero tidal Schwarzschild-like wormhole, at zero temperature. Firstly, we obtain the metric in isotropic coordinates, finding the allowed angular velocities and the circular orbit radii for a material particle as well as for the photon. Following, we compute the changes induced by both gravity and rotation of the plates in the energy density of the quantum vacuum fluctuations associated to the scalar field, in the zero tidal approximation inside the cavity. Finally, the Casimir energy obtained for some these wormholes are graphically compared between themselves and also with those ones related to an Ellis wormhole as well as to a Schwarzschild black hole. With this, the gravito-inertial effects on the quantum vacuum fluctuations analyzed in this work allow to recognize and identify both the geometry and topology of the spacetime associated to each one of these objects.}

\vspace{0.5cm}
\end{abstract}

\renewcommand{\thesection}      {\Roman{section}}
\maketitle
\newpage
\section{INTRODUCTION}
{\large
Some solutions of the Einstein's equations of General Relativity, the so-called wormholes, initially suggested in the works of Flamm \cite{Flamm} and Weyl \cite {Weyl 1, Weyl 2}, represent a tunnel or throat that interconnects two regions of the world \cite {Visser}. The first solutions found (Einstein and Rosen \cite{EinsteineRose}, Wheeler \cite{Wheeler} and Kerr \cite{Ker 1, Ker 2, Ker 3}) were, in general, non-traversable. However, a considerable progress was made when Michael Morris and Kip Thorne, taking into account static, spherically symmetrical and non-rotating wormholes, found a traversable solution \cite{KipThorne}. It is worth remark that the actual increasing research on wormholes has its principal motivation in the discovery of the deep connection exists between these objects and quantum entanglement \cite{Maldacena}.

It is possible to show that the Schwarzschild's solution, by means of an immersion diagram, can configure a wormhole, but it is not traversable, since the edge of its throat coincides with the event horizon. Nevertheless, we can find viable solutions by taking a specific shape for the radial metric component with the redshift function without horizons \cite{KipThorne, MorrisThorne}. An interesting generalization in this sense was carried out in \cite{Cataldo} by taking a linear shape function, the so-called zero tilde Schwarzschild-like wormhole, being static and traversable. Another wormhole based on Schwarzschild black hole solution was obtained by Damour and Solodukhin \cite{Damour}, upon modifying the redshift function adding to it a very tiny positive constant.

In general, for the stability of a wormhole, it is necessary that there is a kind of exotic matter around, a hypothesis supported by the observation of the accelerated expansion of the universe and the suggestion of the dark energy as possible explanation. But in some cases this restriction is unnecessary \cite{Bronnikov, Richarte, Eiroa}. Fluctuations in the vacuum state of quantum fields, which generally have negative energy, can also play a decisive role in defining certain characteristics of the wormholes, such as allowing the existence of closed timelike curves \cite{Kim}. 

The best known physical manifestation of the quantum vacuum fluctuations, the Casimir \cite{Casimir} effect, which has been explored in different contexts \cite{trabalhocasi1,trabalhocasi2,trabalhocasi3,trabalhocasi4,Wilson}, was originally related to the force that arises between two neutral and parallel metallic conductors - in general, plates - in an ideal vacuum of the Minkowsky's spacetime. This force is associated with a quantum effect due to changes in the zero point oscillations of the electromagnetic field as a result of the existence of material boundaries. The Casimir effect can also be associated with cavities immersed in the vacuum of a more general spacetime or with the geometry and topology itself \cite{Mota}. In this sense, several works in the literature have tried to describe this phenomenon in space-times associated with wormholes \cite{wormholecasimir1,wormholecasimir2,wormholecasimir3,wormholecasimir4}. In particular, Sorge \cite{Sorge} recently demonstrated a change in the energy density of the vacuum between two plates which orbit an Ellis wormhole \cite{Ellis}. Such a change would occur due to the geometry of spacetime as well as to the inertial effects coming from rotation of the Casimir apparatus around the wormhole.

Therefore, in this article we will seek to identify, by using the Sorge \cite{Sorge} method, changes of gravito-inertial nature in the quantum vacuum energy density of a massless scalar field confined inside a Casimir cavity orbiting a static and zero tidal Schwarzschild-like wormhole, which is a particular case considered in \cite{Teo,Krasnikov}. We will also analyze the conditions for the existence of the circular orbits in which the plates could travel. This study will permit indirectly distinguish, by means of the Casimir effect measurements, the several astrophysical objects (namely, wormholes and black hole) eventually orbited by a cavity of the type considered here. In a near future, such an experiment could in fact be adaptable for analogues of wormholes built from condensed matter systems, as those ones described in recent works \cite{Allan,Herrero,Flayac,Cedric}. Thus, this work explores a new and different context in which the Casimir effect appears, being necessary in order to open way to a more realistic and complete understanding of the involved phenomena.

The work is divided as follows: In section II, we will find the zero tidal Schwarzschild-like wormhole metric in isotropic coordinates and compute the angular velocities as well as the orbit radii allowed for the Casimir cavity. In section III we will calculate the changes in the energy density of the quantum vacuum fluctuations associated to the massless scalar field inside the cavity orbiting the wormhole, comparing it with those ones related to the Ellis wormhole and to the Schwarzschild black hole. Finally, in section IV we will conclude and close the paper.

\section{SCHWARZSCHILD-LIKE WORMHOLE AND THE ORBITS AROUND IT}

Initially, we must obtain the isotropic form of the traversable wormhole under consideration, which in Schwarzschild coordinates is given by \cite{Cataldo}
\begin{equation}\label{SLW-metric}
ds^2=dt^2-\frac{d\rho^2}{(1-\beta)\left(1-\frac{b_0}{\rho}\right)}-\rho^2(d\theta^2+\sin^2{\theta}d\phi^2),
\end{equation}
where $b_0$ is the wormhole throat radius and $\beta$ is a free parameter of the model, which characterizes the object. The  solution with $\beta=0$ is a specific example of the wormhole firstly found by Teo \cite{Teo}, when its rotational parameter is turned off, and $\beta=-2$ characterizes a Schwarzschild-like wormhole that is a particular Birkhoff solution obtained in the context of a bumblebee gravity model with Lorentz symmetry violation \cite{Casana}.\\
\indent Our objective is initially to find the corresponding solution in isotropic coordinates, such that we have the following spatial line element
\begin{equation}\label{Line Spatial Element}
d\sigma^2=[F(r)]^2[dr^2+r^2 (d\theta^2+\sin^2{\theta}d\phi^2)]=\frac{d\rho^2}{(1-\beta)\left(1-\frac{b_0}{\rho}\right)}+\rho^2 (d\theta^2+\sin^2{\theta}d\phi^2),
\end{equation}
from which we immediately infer $\rho=rF(r)$. Deriving this, we get $\rho'=F(r)+rF'(r)$. On the other hand, we have, by simple comparison of the line elements in Eq. (\ref{Line Spatial Element}),
\begin{eqnarray}
    F(r)dr=\frac{d\rho}{\sqrt{(1-\beta)\left(1-\frac{b_0}{\rho}\right)}}\Rightarrow \rho'=F(r)\sqrt{(1-\beta)\left(1-\frac{b_0}{\rho}\right)}.
\end{eqnarray}
Thus, the conform factor function $F(r)$ can be obtained by solving the ODE
\begin{equation}\label{ODE}
F(r)\sqrt{(1-\beta)\left[1-\frac{b_0}{rF(r)}\right]}=r\frac{dF(r)}{dr}+F(r),
\end{equation}
which yields
\begin{equation}\label{conformal}
F(r)=\frac{(r/b_0)^{\sqrt{1-\beta}-1}}{1-\beta}\left[1+\frac{(1-\beta)}{4(r/b_0)^{\sqrt{1-\beta}}}\right]^2,
\end{equation}
where the integration constant was found in such a way that one obtains the spatial part of the isotropic Schwarzschild solution when $\beta=0$, upon making $b_0=2M$ and observing the correct dimensions of the involved quantities. Hence we obtain
\begin{equation}
ds^2=dt^2-\frac{(r/b_0)^{2(\sqrt{1-\beta}-1)}}{(1-\beta)^2}\left[1+\frac{(1-\beta)}{4(r/b_0)^{\sqrt{1-\beta}}}\right]^4[dr^2+r^2 (d\theta^2+\sin^2{\theta}d\phi^2)],
\end{equation}
\begin{figure}[H]
\centering
\includegraphics[width=0.65\textwidth]{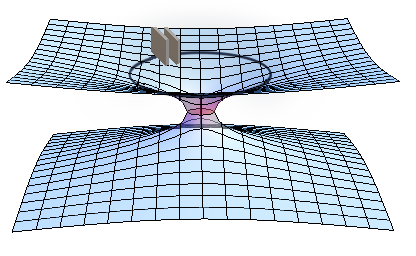}
\caption{The Casimir apparatus orbiting a wormhole.}\label{f1}
\end{figure} 
\noindent which is the static and zero tidal Schwarzschild-like wormhole metric in isotropic coordinates. In this form, it becomes easier introducing the parallel plates geometry of the Casimir apparatus via Cartesian coordinates.

\indent In what follows we will investigate, adopting Sorge's approach \cite{Sorge}, the conditions under which the Casimir apparatus can travel through a circular trajectory around the Schwarzschild-like wormhole, at the equatorial plane ($\theta=\pi/2$). For this, we initially introduce the unit tangent timelike vector ${\bf u}=e^{\psi}(\partial_t+\Omega\partial_\phi)$, where $\Omega=d\phi/dt$ is the angular velocity of the plates and
\begin{equation}
\displaystyle e^\psi   = \frac{1}{\sqrt{1-r^2\Omega^2 F^2(r)}}.  
\end{equation}
Hence we infer the restriction
\begin{equation}
    0\leq \Omega<\displaystyle \frac{(1-\beta)}{b_0(r/b_0)^{\sqrt{1-\beta}}}\left[1+\frac{(1-\beta)}{4(r/b_0)^{\sqrt{1-\beta}}}\right]^{-2}\equiv \Omega(r),
\end{equation} 
keeping only the positive direction of rotation.\\
\indent Introducing the energy and angular momentum per mass unit, $\gamma =E/m$ and $\lambda=L/m$, respectively, which are constants of motion, we get
\begin{equation}
 \displaystyle \frac{dt}{d\tau}=\gamma, \;\;\;\frac{d\phi}{d\tau}=\frac{\lambda}{r^2F^2(r)},
\end{equation}
which lead us to
\begin{equation}
  \displaystyle\lambda=\gamma\Omega r^2F^2(r).\label{lambda}
\end{equation}
The radial equation can be written as 
\begin{equation}\label{radial}
  \displaystyle  \left(\frac{dr}{d\tau}\right)^2=\frac{1}{F^2(r)}\left[\gamma^2-1-\frac{\lambda^2}{r^2F^2(r)}\right],
\end{equation}
and the condition to circular orbits ($dr/d\tau=0$) yields
\begin{equation}\label{lambda2}
   \displaystyle \lambda^2=({\gamma}^2-1)r^2F^2(r),
\end{equation}
and then we have the following angular velocities for a material particle
\begin{equation}\label{Omega}
\displaystyle \Omega(r,\gamma)= \frac{\sqrt{\gamma^2-1}}{\gamma}\frac{(1-\beta)}{b_0(r/b_0)^{\sqrt{1-\beta}}}\left[1+\frac{(1-\beta)}{4(r/b_0)^{\sqrt{1-\beta}}}\right]^{-2}\leq \Omega_{max}.
\end{equation}
We can notice that $\Omega_{max}$, the greatest permitted angular velocity, is exactly the one of massless particles ($m\to 0, \gamma\to \infty $). From Eqs. (\ref{lambda}), (\ref{lambda2}), (\ref{Omega}), and using the expression for $F(r)$ given in Eq. (\ref{conformal}), we arrive at the radii of the circular trajectories
\begin{equation}
   \displaystyle r=b_0\left(\frac{(1-\beta)}{2}\left\{\left(\frac{\lambda}{b_0\sqrt{\gamma^2-1}}-\frac{1}{2}\right)\pm\sqrt{\frac{\lambda}{b_0}\left[\frac{\lambda}{b_0(\gamma^2-1)}-\frac{1}{\sqrt{\gamma-1}}\right]} \right\}\right)^{\frac{1}{\sqrt{(1-\beta)}}},\label{eq13}
\end{equation}
and hence we conclude that there exists circular orbits when $\displaystyle \lambda\geq b_0 \sqrt{\gamma^2-1}$.

In order to find the light ray radii we make $\gamma\rightarrow\infty$  in Eq. (\ref{eq13}), obtaining
\begin{equation}
    \displaystyle r=b_0\left(\frac{1-\beta}{4}\right)^{\frac{1}{\sqrt{1-\beta}}}(-1)^{\frac{1}{\sqrt{1-\beta}}}.
\end{equation}
The expression above will be real and positive if  $\displaystyle \frac{1}{\sqrt{1-\beta}}=2n$, for all integer $n$. Thus we have that only some wormholes hold photon spheres around them, which have space-times with parameters $\beta_n$ given by
\begin{equation}
    \displaystyle  \beta_n=1-\frac{1}{4n^2}.
\end{equation}
With this, the radii of the photon spheres are
\begin{equation}
  r_n=b_0\left(\frac{1}{4n}\right)^{4n}.
\end{equation}
We can notice that these radii are always smaller than the wormhole mouth radius, $b_0$. They would have the same size in the extreme case $n=0$ (or $\beta\to\infty$). 

From Eq. (\ref{radial}), the effective potential per mass unit for circular trajectories is given by
\begin{equation}
 \displaystyle   V(r)=\sqrt{1+\frac{\lambda^2}{r^2F^2(r)}}.
\end{equation}
A critical point of $V(r)$ is
\begin{equation}
  \displaystyle \bar{r}=b_0\left(\frac{1-\beta}{4}\right)^{\frac{1}{\sqrt{1-\beta}}},
\end{equation}
and $\partial^2 V(\bar{r})/\partial r^2<0$. Thus, $\bar{r}$ is a local maximum of $V(r)$ and, consequently, the circular geodesic orbit is unstable. We notice that this trajectory will stay at and out of the wormhole throat, $r\geq b_0$, for $\beta\leq -3$.

From the last equation, we can find the largest radius of the (unstable) geodesic circular trajectory, which is given by the curious expression
\begin{equation}
   r_{max}=b_0e^{1/e}\approx 1.44 b_0,
\end{equation}
which belongs to the wormhole with the particular parameter $\beta=1-4e^2\approx -28.5$.

\section{THE CASIMIR ENERGY}
Provided the possible circular trajectories of the Casimir cavity, we will consider now the quantum vacuum fluctuations of a massless scalar field $\varphi(x^{\mu})$ confined into the orbiting device. Following once more the Sorge's approach, we must initially solve the Klein–Gordon equation, assuming Dirichlet boundary conditions at the cavity walls, whose proper separation, in the comoving observer’s frame, is $L$. The area of each plate is $S$ and we will work with the approximation $L \ll \sqrt{S} \ll b_0 \leq r$. Thus, with this approach we do not consider tidal effects inside the cavity, only the ones of gravito-inertial nature. This latter are taken into account when one implements a rotational frame associated to the apparatus, so that the azimuth angle transforms as $d\phi\to d\phi+\Omega dt$. Thus, the comoving observer will measure a novel metric given by
\begin{equation}\label{rotating metric}
    ds^2=\left[1-r^2\Omega^2 F^2(r)\right] dt^2- F^2(r)(dr^2+r^2d\theta^2-r^2d\phi^2-2\Omega r^2d\phi dt),
\end{equation}
where $F(r)$ is given by Eq. (\ref{conformal}). In order to consider the rectangular geometry of the plates, we will introduce an orthonormal vierbein frame ${\bf \hat{e}}_{\mu}$ given by
\begin{eqnarray}
    &{\bf \hat{e}}_{\tau}&=[1-r^2\Omega^2 F^2(r)]^{-1/2}\frac{\partial}{\partial t},\nonumber\\
    &{\bf \hat{e}}_{x}&=F(r)^{-1}\frac{\partial}{\partial r},\nonumber\\
    &{\bf \hat{e}}_{y}&=[r F(r)]^{-1}\frac{\partial}{\partial \theta},\nonumber\\
    &{\bf \hat{e}}_{z}&=r\Omega F(r)[1-r^2\Omega^2 F^2(r)]^{-1/2}\frac{\partial}{\partial t}+[r F(r)]^{-1}[1-r^2\Omega^2 F^2(r)]^{1/2}\frac{\partial}{\partial \phi},\label{frame}
\end{eqnarray}
where the coordinate $z$ is oriented in the perpendicular direction to the plates, placed at the equatorial plane ($\theta=\pi/2$), and  ${\bf \hat{e}}_{\tau}$ defines the 4-velocity of the static comoving observer.  

The Klein-Gordon equation  in general coordinates, is given by 
\begin{equation}
    \frac{1}{\sqrt{-\hat{g}}}\partial_{\mu}[\sqrt{-\hat{g}}\hat{g}^{\mu\nu}\partial_{\nu}\varphi(x)]+\xi R(x)\varphi(x)=0\label{klein-gordon-geral},
\end{equation}
where $R(x)$ is the Ricci scalar, $\xi$ is a coupling constant, $\hat{g}^{\mu\nu}$ is the inverse of the metric tensor given by $\hat{g}_{\mu\nu}=\hat{{\bf e}}_{\mu}\cdot \hat{{\bf e}}_{\nu}$ and $\hat{g}$ the determinant of this latter. Applying to the Eq. (\ref{klein-gordon-geral}) the vierbein frame for the massless scalar field, considering the minimal coupling  prescription ($\xi=0$), we get
\begin{equation}\label{K-G}
 \nabla^2\varphi-\frac{2}{r}\frac{\partial \varphi}{\partial x}=0,
\end{equation}
where $\nabla^2=\partial^2_x+\partial^2_y+\partial^2_z$. We have used the approximation of zero tidal inside the cavity ($r\approx$ constant, where the plates are placed).

The solutions of Eq. (\ref{K-G}) satisfying the Dirichlet boundary conditions are given by
\begin{equation}
\varphi_{n,{\bf k_{\|}}}=N_n\exp{({-i\omega_{n,{\bf k_{\|}}}}\tau)}\exp{(i{\bf k_{\|}}}\cdot {\bf x_{\|}})\sin{\left(\frac{n\pi z}{L}\right)},
\end{equation}
where ($\omega_{n,\|},{\bf k_{\|}})$ are the eigenfrequencies and momenta of the field propagation modes parallel to the plates. The modes are normalized from the Klein-Gordon scalar product, given in the vierbein frame by \cite{Sorge}
\begin{equation}
\displaystyle \langle \varphi_n (\textbf{k}), \varphi_m(\textbf{k}')\rangle = i \int_\sum [(\partial_a \varphi_n){\varphi^*_m} - \varphi_n(\partial_a{\varphi^*_m})]n^a dxdydz, 
\end{equation}
where $n^a = e^a_\mu n^\mu$ and $n^\mu = (1, 0, 0, - \Omega)$. Taking into account that
\begin{equation}
\displaystyle\langle \varphi_n(\textbf{k}_\parallel), \varphi_m(\textbf{k}'_\parallel) \rangle = \delta^2 (\textbf{k}_\parallel - \textbf{k}'_\parallel) \delta_{mn},    
\end{equation}
we arrive at the normalization parameter
\begin{equation}
    N_n=\left(\frac{\sqrt{1-r^2\Omega^2F^2(r)}}{4\pi^2L\omega_{n,\|}}\right)^{1/2},
\end{equation}
 which guarantees the orthonormalization of the field modes and encodes the properties of the spacetime under inspection.

The Casimir energy will be obtained from the regularization of the expected value of the quantum vacuum fluctuations energy, given by \begin{equation} \label{MeanEnergy}
    \langle \epsilon \rangle=\frac{1}{V_p}\int_{\Sigma}d^3{\bf x}\sqrt{g_{\Sigma}} \sum_n\int d^2{\bf k_{\|}}T_{00},
\end{equation}
where the first integration is realized in the Casimir cavity, which has proper volume $V_p=V \sqrt{-g_{\Sigma}}$, with $V$ being the  volume measured by a distant observer, $g_{\Sigma}=\det{(\hat{g}_{\mu\nu})/\hat{g}_{tt}}$ \cite{Birrel,Zhang}, and the second integration is over the space of the momenta parallel to the plates. $T_{00}$ is the purely temporal component of the energy-momentum tensor, given by
\begin{equation}\label{MomentEnergy}
    T_{00}=\partial_{\tau}\varphi_n\partial_{\tau}\varphi_n^{*}-\frac{1}{2}\eta^{ij}\partial_i\varphi_n\partial_j\varphi_n^{*}.
\end{equation} 
Plugging (\ref{MomentEnergy}) into (\ref{MeanEnergy}), we obtain
\begin{equation}
    \langle \epsilon \rangle=\frac{\sqrt{1-r^2\Omega^2F^2(r)}}{8\pi^2L}\sum_n\int_0^{\infty} d^2\mathbf{k}_{{\|}}\sqrt{k^2_{\|}+\frac{n^2\pi^2}{L^2}}.\end{equation}
From the Schwinger proper-time representation for the above integral, given by
\begin{equation}
    a^{-z}=\frac{1}{\Gamma(z)}\int_0^{\infty}t^{z-1}\exp{(-a t)}dt,
\end{equation}
in which $a=k^2+n^2\pi^2/L^2$ and $z=-1/2$, with $k_{\|}=k$ and $d^2k_{\|}=2\pi k dk$, after firstly performing the easy integral in momentum variable, the remaining integral in $t$ can be made using the Euler Representation for the gamma function, which is
\begin{equation}
    \Gamma(w)=\int_0^{\infty}x^{w-1}\exp{(-x)}dx,
\end{equation}
and the summation in $n$ is carried out employing the definition of the Riemann zeta function, $\zeta(s)=\sum_1^{\infty}n^{-s}$. With this, we finally arrive at the Casimir energy density between the plates, given by
\begin{equation}
 \epsilon_C =-\sqrt{1-r^2\Omega^2\frac{(r/b_0)^{2(\sqrt{1-\beta}-1)}}{(1-\beta)^2}\left[1+\frac{(1-\beta)}{4(r/b_0)^{\sqrt{1-\beta}}}\right]^4}\frac{\pi^2}{1440L^4},
\end{equation}
where we taken into account the expression for $F(r)$ given in Eq. (\ref{conformal}). We notice that the negative of the multiplicative factor out of the squared root is the Casimir energy density ($\epsilon_0$) of the scalar field inside the plates situated in the Minkowsky spacetime. Thus, the modulus of the found quantity is always lower than this latter.

In the graph of Figure 2, we have depicted the ratio  $R=\epsilon_{C}/|\epsilon_0|$ as a function of the radial coordinate, for $b_0=1$, $\Omega=0.2$, and some values of $\beta$. The angular velocity $\Omega$ was chosen between the permitted ones. We also depicted the Casimir energy density of the Ellis wormhole according to \cite{Sorge}, as well as the one related to the Schwarzschild black hole, given in \cite{Sorge2} upon vanishing the angular momentum of the Kerr black hole, using isotropic coordinates and adopting the same parameters where they are applicable. In this latter case, $b_0=2M$.
\begin{figure}[H]
\centering
\includegraphics[scale=0.4]{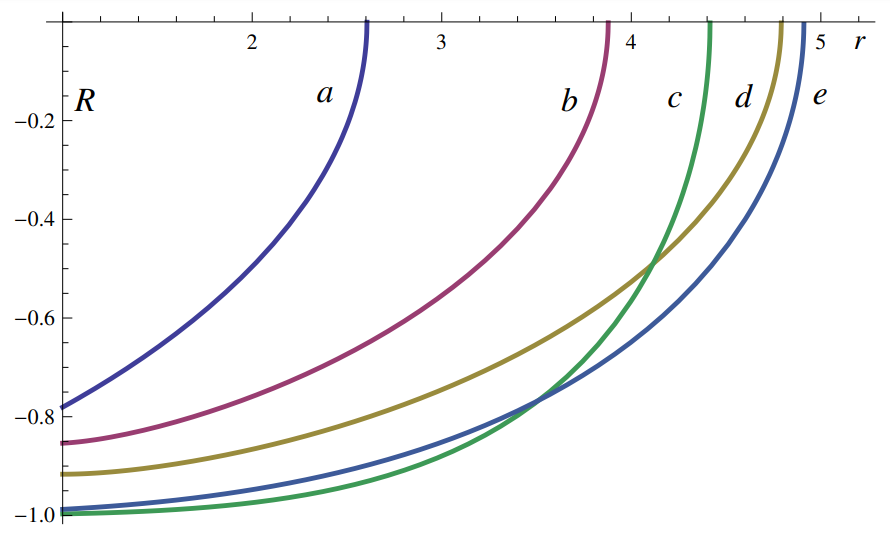}
\caption{The ratio $R=\epsilon_{C}/|\epsilon_0|$, as a function of the radial coordinate, $r>b_0$, for $b_0=1$, $\Omega=0.2$. a) Schwarzschild-like (S-L) wormhole with $\beta=0.5$. b) Schwarzschild black hole. c) S-L wormhole with $\beta=-3$. d) Ellis wormhole. e) S-L wormhole with $\beta=-1$.}
\end{figure}

At the wormhole throat $r=b_0$, the Casimir energy density is given by
\begin{equation}
  \epsilon_C=  -\sqrt{1-\frac{b_0^2\Omega^2}{(1-\beta)^2}\left[1+\frac{(1-\beta)}{4}\right]^4}\frac{\pi^2}{1440L^4}.
\end{equation}
In this case, the Casimir energy density is lower than the one measured at the Ellis wormhole throat when $-11-8\sqrt{2}<\beta<8\sqrt{2}-11$, for the same allowed parameters.

\section{FINAL REMARKS}

In this paper, based on Sorge's approach \cite{Sorge} we have investigated the Casimir effect around a static and zero tidal Schwarzschild-like wormhole, in the parallel plates configuration, at zero temperature. Initially, we have obtained the corresponding spacetime metric in isotropic coordinates in order to implement the rectangular geometry of the plates. 

Following, we studied the circular trajectories allowed to a particle ({\it i.e.} the plates) travelling around the object, finding the possible angular velocities and respective orbit radii. Specifically, for massless particles, we verified that only some wormholes with certain values ({\it i.e.}``quantized'') for the parameter $\beta$ permit the existence of photon spheres. We also found the radii of (unstable) geodesic circular trajectories and the wormhole with the largest value among them, for a same throat radius. 

Continuing, we solved the Klein-Gordon equation for a massless scalar field  with minimum coupling in a rotational comoving reference frame, in the above referred metric, imposing Dirichlet boundary conditions on the plates which orbit circularly the wormhole. Here we have used the approximation that neglects tidal effects inside the cavity. Then we obtained the expected value of the energy density of the quantum vacuum fluctuations associated to the confined field. After regularizing  this otherwise divergent quantity, we arrived at the Casimir energy density, which is lower than the one obtained in the Minkowsky spacetime for the same configuration of the plates. Then we have compared graphically this energy density, for some values of $\beta$, with the ones obtained in both the Ellis wormhole and Schwarzschild black hole space-times, according to \cite{Sorge} and \cite{Sorge2}, respectively.  

Though we have considered the zero temperature scenario, the immersion of the plates in a thermal bath will not change the results found in the Minkowski spacetime, according to Zhang reasoning \cite{Zhang}, since the dependence of the Casimir thermodynamical quantities on the proper temperature of the plates, {\it i.e.}, the one measured by a co-moving observer, would have the same form presented in that spacetime.

We conclude that the gravito-inertial effects on vacuum quantum fluctuations studied here allow indirectly detecting both the geometry and topology of the spacetime, enabling the distinction among the possible alternatives, such as different types of wormholes and a black hole. In other words, an experimental curve of eventual measurements of the Casimir energy obtained at different positions with respect to the wormhole (or black hole) center should be superimposed to the theoretical graphics shown in Figure 2, which would allow, in principle, the inference of the object one is facing. As a future perspective, we intend expand this analysis to cosmological scenarios.

\section*{ACKNOWLEDGEMENTS}

CRM would like to thank Conselho Nacional de Desenvolvimento Cient\'{i}fico e Tecnol\'{o}gico (CNPq) and Fundação Cearense de Apoio ao Desenvolvimento Científico e Tecnológico (FUNCAP), under grant PRONEM PNE-0112-00085.01.00/16, for the partial financial support.

\bibliographystyle{unsrt}  


\begin{thebibliography}{1}

\bibitem{Flamm} Flamm, L. Beitrage zur Einsteinschen Gravitationstheorie. Phys. Z. \textbf{17}, 448 (1916).
\bibitem{Weyl 1} Weyl, H. Philosophie der Mathematik und Naturwissenschaft. Handbuch der philosophie. Leibniz Verlag, Munich, (1928).
\bibitem{Weyl 2} Weyl, H. Philosophy of Mathematics and Natural Science. Princeton University Press, Princeton, (1949).
\bibitem{Visser} Visser, M. Lorentzian Wormholes: From Einstein to Hawking. American Institute of Physics, New York, (1996).
\bibitem{EinsteineRose} Einstein, A.; and Rosen, N. The particle problem in the general theory of relativity. Phys. Rev. \textbf{48}, 73 (1935).
\bibitem{Wheeler} Misner, C. W.; and Wheeler, J. A. Classical physics as geometry: gravitation, electromagnetism, unquantized charge, and mass as properties of curved empty space. Ann. Phys. \textbf{2}, 525 (1957).
\bibitem{Ker 1} Hawking, S, W.; and Ellis, G. F. R. The Large Scale Structure of Space- Time. Cambridge University Press, Cambridge, (1973).
\bibitem{Ker 2} Misner, C. W.; Thorne, K. S.; and Wheeler, J. A. Gravitation. W. H. Freeman and Company, San Francisco, (1973).
\bibitem{Ker 3} Waldo, R. M. General Relativity. University of Chicago Press, Chicago, (1984).
\bibitem{KipThorne} Morris, M. S.; and Thorne, K. S. Wormholes in spacetime and their use for interstellar travel: A tool for teaching general relativity. Am. J. Phys. \textbf{56}, 395 (1988).
\bibitem{Maldacena} Maldacena, J and Susskind, L, Cool horizons for entangled black holes, Forts. Phys. {\bf61}, 9, 781 (2013).
\bibitem{MorrisThorne} Morris, M.S.; Thorne, K.S.; and Yurtsever, U. Wormholes, Time Machines, and the Weak Energy Condition. Phys. Rev.Lett. \textbf{61}, 1446 (1988).
\bibitem{Cataldo} Cataldo, M.; Liempi, L.; and Rodriguez, P. Traversable Schwarzschild-like wormholes. Eur.Phys. J. C {\bf77}, 748 (2017). 
\bibitem{Damour} Damour, T.; and Solodukhin, S. N. Wormholes as Black Hole Foils. Phys. Rev. D {\bf 76}, 024016 (2007).
\bibitem{Bronnikov} Bronnikov, K. A.; and Skvortsova, M. V. Cylindrically and axially symmetric wormholes. Throats in vacuum?  Grav. Cosmol. \textbf{20}, 171 (2014).
\bibitem{Richarte} Richarte, M. G. Cylindrical wormholes with positive cosmological constant. Phys. Rev. D \textbf{88}, 027507 (2013).
\bibitem{Eiroa} Eiroa, E. F.; and Simeone, C. Cylindrical thin-shell wormholes. Phys. Rev. D \textbf{82}, 084039 (2010).
\bibitem{Kim} Kim, S.-W.; and Thorne, K. S. Do Vacuum Fluctuations Prevent the Creation of Closed  Timelike Curves? Phys. Rev. D \textbf{43}, 3939 (1991).
\bibitem{Casimir} Casimir, H. B. G. On the attraction between two perfectly conducting plates, Proc. Kon. Ned. Akad. Wet. \textbf{51}, 793 (1948).
\bibitem{trabalhocasi1} Sorge, F. Casimir effect in a weak gravitational field. Class. Quant. Grav. \textbf{22}, 5109 (2005).
\bibitem{trabalhocasi2} Muniz, C. R.; Bezerra, V. B.; and Cunha, M. S. Casimir effect in the Hořava–Lifshitz gravity with a cosmological constant. Ann. Phys. \textbf{359}, 55 (2015).
\bibitem{trabalhocasi3} Bezerra, V. B.; Cunha, M. S.; Freitas, L. F. F.; Muniz, C. R.; and Tahim, M. O. Casimir effect in the Kerr spacetime with quintessence. Mod. Phys. Lett. A \textbf{32}, 1750005 (2017). 
\bibitem{trabalhocasi4} Lima,  A.P.C.M., Alencar, G., Muniz, C.R., and Landim, R.R., Null second order corrections to Casimir energy in weak gravitational field, JCAP 2019, 11 (2019).
\bibitem{Wilson} Wilson, J.H., Sorge, F., and Fulling, S.A., Tidal and nonequilibrium Casimir effects in free fall, Phys. Rev. {\bf D101}, 065007 (2020).
\bibitem{Mota} Bezerra, V. B.; Mota, H. F.; and Muniz, C. R. Remarks on a gravitational analogue of the Casimir effect. Int. J. Mod. Phys. D {\bf 25}, 1641018 (2016).
\bibitem{wormholecasimir1} Garattini, R. Casimir wormholes. Eur. Phys. J. C \textbf{79}, 951 (2019).
\bibitem{wormholecasimir2} Jusufi, K.; Channuie, P.; and Jamil, M. Traversable wormholes supported by GUP corrected Casimir energy. Eur. Phys. J. C \textbf{80}, 127 (2020). 
\bibitem{wormholecasimir3} Khabibullin, A. R.; Khusnutdinov, N. R.; and Sushkov, S. V. The Casimir effect in a wormhole spacetime. Class. Quant. Grav. \textbf{23}, 627 (2006). 
\bibitem{wormholecasimir4} Butcher, L. M. Casimir energy of a long wormhole throat, Phys. Rev. D {\bf 90}, 024019 (2014).
\bibitem{Sorge} Sorge, F. Casimir effect around an Ellis wormhole. Int. J. Mod. Phys. D \textbf{29}, (2019).
\bibitem{Ellis} Ellis, H. G. Ether flow through a drainhole - a particle model in general relativity. J. Math. Phys. \textbf{14}, 104 (1973).
\bibitem{Teo} Teo, E. Rotating traversable wormholes. Phys. Rev. D {\bf 58}, 024014 (1998).
\bibitem{Krasnikov} Krasnikov, S. Schwarzschild-like wormholes as accelerators. Phys. Rev. D. \textbf{98}, (2018).
\bibitem{Allan} Greenleaf, A., Kurylev, Y., Lassas, M., and Uhlmann, G., Electromagnetic Wormholes and Virtual Magnetic Monopoles from Metamaterials, Phys. Rev. Lett. {\bf 99}, 183901 (2007).
\bibitem{Herrero} González, J. and Herrero, J., Graphene wormholes: A condensed matter illustration of Dirac fermions in curved space, Nucl.Phys. {\bf B825}, 3, 426 (2010).
\bibitem{Flayac} Solnyshkov, D.D., Flayac, H., and Malpuech, G., Black holes and wormholes in spinor polariton condensates, Phys.Rev.{\bf B84}, 233405 (2011).
\bibitem{Cedric} Peloquin, C.; Euvé, L.P.; Philbin, T. and Rousseaux, G. Analog wormholes and black hole laser effects in hydrodynamics, Phys. Rev. {\bf D93}, 084032 (2016). 
\bibitem{Casana} Casana, R.; Cavalcante, A.; Poulis, F. P.; and Santos, E. B. An exact Schwarzschild-like solution in a bumblebee gravity model. Phys. Rev. D {\bf 97}, 104001 (2018). 
\bibitem{Birrel} N. D. Birrell and P. C. W. Davies, Quantum Fields in Curved Space, Cambridge University Press, Cambridge, England, (1982).
\bibitem{Zhang} A. Zhang, Thermal Casimir Effect in Kerr Space-time, Nuc.Phys.{\bf B898}, 220 (2015) 
\bibitem{Sorge2} Sorge, F. Casimir energy in Kerr space-time. Phys. Rev. D {\bf 90}, 084050 (2014).



\end{thebibliography}

}

\end{document}